\documentclass[aps,pra,twocolumn,superscriptaddress,showpacs]{revtex4-1}
\usepackage{ulem,bm}
\usepackage{amsmath, upgreek, amssymb, graphicx, paralist, gensymb,epstopdf}
\usepackage[colorlinks, linkcolor=blue, citecolor=blue, urlcolor=blue]{hyperref}
\usepackage{array,epsfig,amsmath,amssymb}
\usepackage{amsmath,epstopdf}
\usepackage{graphicx} 
\usepackage{dsfont}
\usepackage{color}
\newcommand{\ket}[1]{\left\vert#1\right\rangle}
\newcommand{\bra}[1]{\left\langle#1\right\vert}

\def\bra#1{\langle #1|}
\def\ket#1{|#1 \rangle}
\def\bracket#1#2{\langle #1|#2 \rangle}

\begin{document}
\title{Single-photon characteristics of superposed weak coherent states}

\author{Seung-Woo Lee}
\affiliation{Quantum Universe Center, Korea Institute for Advanced Study, Seoul 02455, Korea}

\author{Jaewan Kim}
\affiliation{School of Computational Sciences, Korea Institute for Advanced Study, Seoul 02455, Korea}


\begin{abstract}

We study a superposed weak coherent state that can fundamentally mimic an ideal single photon not only with respect to the number of photons but also in terms of an indeterminate phase. It is close to the single-photon state with almost unit fidelity as well as exhibits fundamental features of single photons such as antibunching and Hong-Ou-Mandel interference. The emergence and vanishing of single-photon characteristics can be directly observed by changing two parameters, i.e.,~mean photon number and number of phases. Our result shows that the uncertainty between photon number and phase indeed constitutes the characteristics of single photons. Finally, we apply the superposed weak coherent state to quantum key distribution and demonstrate that it outperforms the typical approach using phase-randomized weak coherent states.

\end{abstract}

\maketitle

\section{Introduction}

Single photons are ideal carriers of quantum information and primary resources to explore quantum mechanics. Quantum principles such as complementarity and the impossibility of cloning quantum state of a single photon \cite{noclon} constitute the fundamental basis of the security of quantum communications \cite{BB84,Ekert91,Gisin2002,Scarani2009,Lo2014}. Especially, the uncertainty in the measurement of two conjugate variables on single photons is the key feature of quantum key distribution (QKD) \cite{BB84,Gisin2002,Scarani2009}. Moreover, their characteristics such as antibunching and Hong-Ou-Mandel interference \cite{HOM} play a major role in implementing quantum communications \cite{Teleportation,DLCZ,Pirandola2015}. single-photon sources have substantially progressed \cite{Eisaman2011,Aharonovich2016}, although they are still challenging \cite{Schneider2018}.

A natural question that comes to mind is whether single photons can be attained by attenuating lasers, e.g., by dividing the pulse many times via beam splitters. An attenuated laser, however, contains more than single photons with a nonzero probability following a Poisson distribution. It can be described by a weak coherent state (WCS), $\ket{\sqrt{\mu} e^{i\theta}}=e^{-\mu/2}\sum^{\infty}_{n=0}(\sqrt{\mu} e^{i\theta})^n/\sqrt{n!}\ket{n}$, where $\mu$ is the mean photon number, $\theta$ is the phase, and $\ket{n}$ is the photon number state. It indeed distributes the uncertainty (roughly) equally between phase and amplitude so that the phase is not indeterminate. Moreover, it never exhibits the behaviors of single photons, i.e.,~antibunching and Hong-Ou-Mandel interference. Although, in certain applications (e.g.~QKD), the multi-photon issue of WCS can be circumvented by the decoy state method \cite{decoy1,decoy2,decoy3}, it requires a perfect random phase $\theta$ \cite{Lo2007}. 
Recent attempts to yield single-photon behaviors from phase randomized WCS \cite{SPbyWCS1,SPbyWCS2,SPbyWCS3,SPbyWCS4} are indirect observations through data reconstruction (but not the genuine exhibitions) of such behaviors. Therefore, what fundamentally constitutes the characteristics of single photons and how they can be modulated in the quantum optical framework are, in fact, still obscure.

In this paper, we study an optical quantum state that can fundamentally mimic single photons with respect to both the number of photons and indeterminate phase. This state, hereafter referred to as {\em pseudo-single-photon state} (PSP), can be defined as a superposition of WCSs equally distributed on a circle in phase space with a discretized phase $2\pi/d$. Not only is it close to the single-photon state with almost unit fidelity but it also exhibits the fundamental features of single photons such as antibunching and Hong-Ou-Mandel interference. Its single-photon characteristics can be modulated with two parameters $\mu$ and $d$; as either $\mu$ decreases or $d$ increases, PSPs and their behaviors become closer to ideal single photons. This reveals the fact that the uncertainty between the photon number and phase indeed constitutes the characteristics of single photons. We also propose a scheme to generate PSPs at room temperature with fast response time employing atomic vapor in hollow-core photonic crystal fibers \cite{KerrOF2,KerrOF3}, which is feasible and suitable for fiber-optic communication. Finally, we show that the single-photon characteristics of PSPs allows us to outperform the typical approach using phase randomized WCSs in QKD. This is a representative example showing the quantum advantage of uncertainty over the statistical randomization.

\section{Pseudo number states} 

We start with the definition of pseudo-number states. In contrast to ideal photon number states having continuous random phase, we consider discretized phases $2\pi/d$ with an integer $d$, corresponding to equally distributed coherent states on a circle in phase space, $\{|\sqrt{\mu}\rangle, |\sqrt{\mu}\omega\rangle,...,|\sqrt{\mu}\omega^{d-1}\rangle\}$ where $\omega=\exp{(2\pi i/d)}$. We define pseudo-number states as the maximally superposed states of the coherent states
\begin{equation}
\label{eq:pns1} 
\ket{j_d}=\frac{1}{\sqrt{{\cal N}_{\mu,j}}}\sum^{d-1}_{q=0}\omega^{-jq}\ket{\sqrt{\mu}\omega^{q+\delta}}
\end{equation}
where $j\in\{0,1,...,d-1\}$, and $\delta$ is the reference phase (hereafter $\delta=0$ unless otherwise necessary). Note that it can be rewritten again by $\ket{j_d}=(d/\sqrt{{\cal N}_{\mu,j}})e^{-\mu/2}\sum^{\infty}_{n\in \mathbb{S}_j}\mu^{n/2}/\sqrt{n!}\ket{n}$, where $\mathbb{S}_j=\{n| j\equiv n(\rm{mod~}d)\}$ as studied in Ref.~\cite{old1,old2,Jkim2015}. The normalization factor can be obtained as ${\cal N}_{\mu,j}=\sum_{q,q'=0}^{d-1} \omega^{j(q'-q)} \bracket{\sqrt{\mu}\omega^{q'}}{\sqrt{\mu}\omega^{q}}= d^2 e^{-\mu}\sum^{\infty}_{n\in \mathbb{S}_j}\mu^{n}/n!$. The coherent state on a circle is written conversely with the pseudo-number states as $\ket{\sqrt{\mu}\omega^q}=(1/\sqrt{d})\sum^{d-1}_{j=0}\omega^{qj}\sqrt{{\cal N}_{\mu,j}/d}\ket{j_d}$. We can see that $\ket{j_d}$ and $\ket{\sqrt{\mu}\omega^q}$ represent conjugate variables connected by the number-phase uncertainty.

For large $\mu$ (in the regime $\sqrt{\mu}>d$), $\ket{j_d}$ nearly constitute $d$-dimensional orthonormal basis \cite{Jkim2015}. As ${\cal N}_{\mu,j}\rightarrow d$, $\ket{j_d}$ and $\ket{\sqrt{\mu}\omega^q}$ are in the discrete Fourier transform relations: $\ket{\sqrt{\mu}\omega^l}=\hat{F}_d\ket{j_d}$ and $\ket{k_d}=\hat{F}^{-1}_d\ket{\sqrt{\mu}\omega^j}$, where $\hat{F}_d=(1/\sqrt{d})\sum^{d-1}_{k=0}\sum^{d-1}_{l=0}\omega^{kl}\ket{k_d}\bra{l_d}$. 

\begin{figure}
\centering
\includegraphics[width=3.4in]{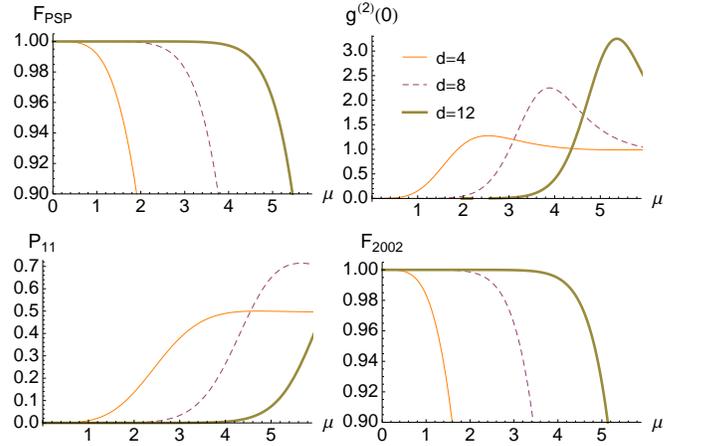}
\caption{$F_{\rm PSP}$: the fidelities between ideal single-photon state and PSPs,  $g^{(2)}(0)$: the second-order correlation function of PSPs, $P_{11}$: the probabilities for detecting photons at both outputs of beam splitter when two PSPs enter different input modes, $F_{2002}$: the fidelities between the output state of the beam splitter and $(\ket{2}\ket{0}-\ket{0}\ket{2})\sqrt{2}$, by changing $\mu$ for $d=4,8,12$.}
\label{fig:FPNS}
\end{figure}

On the other hand, in the small $\mu$ regime, $\ket{j_d}$ can mimic ideal number states, as they become closer to $\ket{j}$ as $\mu$ gets smaller or $d$ increases. The fidelity between $\ket{j_d}$ and $\ket{j}$ is given by
\begin{equation}
\label{eq:fidelitytoideal}
F(\ket{j},\ket{j_d})=|\bracket{j}{j_d}|^2=\bigg|\frac{d}{\sqrt{{\cal N}_{\mu,j}}}e^{-\mu/2}\frac{\mu^{j/2}}{\sqrt{j!}}\bigg|^2.
\end{equation}
For example, the fidelity between ideal and PSP states
\begin{eqnarray}
F_{\rm PSP}\equiv F(\ket{1},\ket{1_d})=\frac{d^2e^{-\mu}\mu}{{\cal N}_{\mu,1}}=\frac{\mu}{\sum^{\infty}_{n\in \mathbb{S}_1}\mu^{n}/n!},
\end{eqnarray}
is plotted in Fig.~\ref{fig:FPNS}. It shows that the fidelity approaches to unit as $\mu$ decreases or $d$ increases. We can analytically see $F_{\rm PSP}=\mu/\mu=1$ in the limit $d\rightarrow\infty$, and  
\begin{eqnarray}
F_{\rm PSP}=\frac{\mu}{\mu+\frac{\mu^{d+1}}{(d+1)!}+\frac{\mu^{2d+1}}{(2d+1)!}+...}\rightarrow1
\end{eqnarray}
in the limit $\mu\rightarrow0$ with finite $d$. We thus hereafter call $\ket{1_d}$ with arbitrarily small $\mu$ PSP state. 

For the case when $d=2$, the PSP becomes the so called odd cat state \cite{OddCat} that is well known to approach $\ket{1}$ when $\mu\rightarrow0$. Note that, compared to the odd cat states, PSPs with higher $d$ get closer to single photons in the wider range of $\mu$.

We can also analyze the effects of photon losses on PSPs. Their evolution in a lossy environment can be evaluated by solving the master equation \cite{Phoenix90},
\begin{equation}
\frac{d\rho}{dt}=\hat{J}\rho+\hat{L}\rho,
\end{equation}
where $\hat{J}\rho=\gamma\hat{a}\rho\hat{a}^{\dag}$ and
$\hat{L}\rho=-\gamma(\hat{a}^{\dag}\hat{a}\rho+\rho\hat{a}^{\dag}\hat{a})/2$ with the annihilation (creation) operator $\hat{a} (\hat{a}^{\dag})$. Here, $\gamma$ is the decay constant, and $\eta=e^{-\gamma t}$ is defined as the transmission rate under loss. The pseudo-number state $\ket{j_d}$ evolves under losses into
\begin{equation}
\begin{aligned}
\rho_j(t)=\frac{1}{{\cal N}_{\mu,j}}\sum^{d-1}_{q,q'=0}&\omega^{j(q'-q)}e^{(\omega^{q-q'}-1)\mu(1-\eta)}\\
&\hspace{10mm}\times\ket{\sqrt{\mu\eta}\omega^q}\bra{\sqrt{\mu\eta}\omega^{q'}}.
\end{aligned}
\end{equation}
For PSP with small $\mu(1-\eta)$, it can be written by 
\begin{equation}
\label{eq:singleLoss}
\begin{aligned}
\rho_1(t)&=\frac{1}{{\cal N}_{\mu,1}}\sum^{d-1}_{q,q'=0}\omega^{q'-q}e^{(\omega^{q-q'}-1)\mu(1-\eta)}\\
&\hspace{32mm}\times\ket{\sqrt{\mu\eta}\omega^q}\bra{\sqrt{\mu\eta}\omega^{q'}}\\
&\approx\big\{1-\mu(1-\eta)\big\}\ket{1'_d}\bra{1'_d}+\mu(1-\eta)\ket{0'_d}\bra{0'_d}
\end{aligned}
\end{equation} 
where $\ket{1'_d}$ and $\ket{0'_d}$ are respectively the pseudo-single and -vacuum states with an attenuated mean photon number $\mu'=\mu\eta$.

\section{single-photon behaviors}

We investigate the single-photon characteristics of PSPs by changing $d$ and $\mu$:

\subsection{Antibunching}

Antibunching refers to sub-Poissonian statistics of photons \cite{antib}, characterized by the second-order correlations function
\begin{eqnarray}
g^{(2)}(0)=\frac{\langle a^{\dag}a^{\dag}aa\rangle}{\langle a^{\dag}a \rangle^2}=\frac{\langle a^{\dag}aa^{\dag}a\rangle-\langle a^{\dag}a \rangle}{\langle a^{\dag}a \rangle^2}<1.
\end{eqnarray}
Note that coherent states are in Poissonian statistics with $g^{(2)}(0)=1$. The statistics with $g^{(2)}(0)>1$ is called super-Poissonian. The second-order correlation function of PSPs can be calculated as
\begin{eqnarray}
g^{(2)}(0)={\cal N}_{\mu,1}\frac{\sum^{d-1}_{q,q'=0}\omega^{q'-q}e^{(\omega^{q-q'}-1)\mu}}{(\sum^{d-1}_{q,q'=0}e^{(\omega^{q-q'}-1)\mu})^2},
\end{eqnarray}
for arbitrary given $\mu$ and $d$ and plotted as in Fig.~\ref{fig:FPNS}. The sub-Poissonian statistics with nearly $g^{(2)}(0)=0$ are observed in small $\mu$ region that becomes wider as $d$ increases. Note that, as $\mu$ increases, first a transition occurs from sub- to super-Poissonian, and then the statistics gradually become Poissonian $g^{(2)}(0)=1$. A similar observation was reported in Ref.~\cite{KSLee94}.

\subsection{Two-photon (Hong-Ou-Mandel) interference}
Assume that two PSPs, with arbitrary ($\mu$, $d$, $\delta$) and ($\mu'$, $d'$, $\delta '$) respectively, enter different input modes of a beam splitter. The state in the two output mode can then be written as
\begin{eqnarray}
\label{eq:BSout}
\begin{aligned}
&\ket{1_d}\ket{1_d'}\xrightarrow{BS}\ket{\rm{out}}=\frac{1}{\sqrt{{\cal N}_{\mu,1}{\cal N}_{\mu',1}}}\sum^{d-1}_{q=0}\sum^{d-1}_{q'=0}\omega_d^{-q-\delta}\omega_{d'}^{-q'-\delta'}\\
&\hspace{8mm}\times\Big|\frac{\sqrt{\mu}\omega_d^{q+\delta}+\sqrt{\mu'}\omega_{d'}^{q'+\delta'}}{\sqrt{2}}\Big\rangle\Big|\frac{\sqrt{\mu}\omega_d^{q+\delta}-\sqrt{\mu'}\omega_{d'}^{q'+\delta'}}{\sqrt{2}}\Big\rangle,
\end{aligned}
\end{eqnarray}
where we use subscript in $\omega_d=\exp(2\pi i/d)$ to distinguish different $d$. The probability for detecting photons at both output modes, $P_{11}$, can be calculated by $P_{11}(\mu,d,\delta|\mu',d',\delta')=\big\langle \rm{out}\big|(\openone-\ket{0}\bra{0})\otimes(\openone-\ket{0}\bra{|0})\big|out\big\rangle$. The fidelity between the state in output modes and $(\ket{2}\ket{0}-\ket{0}\ket{2})/\sqrt{2}$, referred to as $F_{2002}$, is also calculated by $F_{2002}(\mu,d,\delta|\mu',d',\delta')=\big| \bra{\rm{out}} (\ket{2}\ket{0}-\ket{0}\ket{2})/\sqrt{2}\big|^2$. 

We first analyze the case when two identical PSPs enter the input modes, i.e.~$\mu=\mu'$, $d=d'$, and $\delta=\delta'=0$, and plot $P_{11}$ and $F_{2002}$ by changing $\mu$ with different $d$ in Fig.~\ref{fig:FPNS}. We can see that two-photon interference occurs and becomes more evident with increasing $d$ or decreasing $\mu$. This is stark contrast to phase randomized WCSs that cannot exhibit such an interference even with $\mu\rightarrow0$. Compared to odd cat states (PSP with $d=2$) whose $F_{2002}$ is close to 1 only with $\mu\rightarrow0$, PSPs with higher $d$ (for example, with $d=8$) show clear interference with a wider range of $\mu$ ($\mu\lesssim2$) as shown in Fig.~\ref{fig:FPNS}.

In general, two different PSPs (with different $\mu$, $d$, or $\delta$) can exhibit two-photon interference that becomes clearer as the overall mean photon number decreases or the overall number of phases increases. Note that the maximum interference can be observed between the same PSPs when the overall number of photons and phases are fixed. This manifests the fact that maximum interference occurs between the most indistinguishable PSPs. 

\section{A generation scheme}
\label{sec:generation}

\begin{figure}
\centering
\includegraphics[width=3.4in]{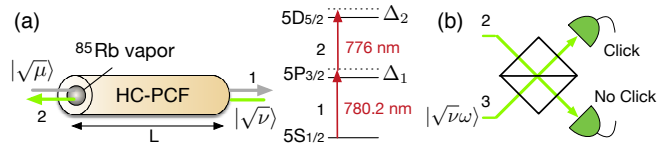}
\caption{(a) Left: Cross-phase modulation with $^{85}$Rb vapor filled HC-PCF can be used as a platform for generating PSPs. Right: $^{85}$Rb energy level. (b) Discrimination of PSP from other pseudo-number states by photon on-off detectors. The event in which only the upper detector clicks guarantees the generation of a PSP in output mode 1, referred to as a triggered event.
}\label{fig:Scheme}
\end{figure}

Let us consider an exemplary scheme to generate PSPs using a cross-phase modulation technique. 

\subsection{Cross-phase modulation with atomic vapor in a hollow-core photonic crystal fiber}

We first consider a feasible scheme of cross-phase modulation, aiming to apply for generating PSPs: Atomic vapor filled in hollow-core photonic crystal fiber (HC-PCF) is an efficient platform for photon-level quantum processors \cite{KerrOF2,KerrOF3,KerrOF4,KerrOF5,KerrOF6,KerrOF7,KerrOF8} such as cross-phase modulation \cite{KerrOF2,KerrOF3}, all-optical switches \cite{KerrOF4,KerrOF5}, and quantum memories \cite{KerrOF7}. 

Particularly, cross-phase modulation has been realized with rubidium (Rb) atoms confined in HC-PCF \cite{KerrOF2,KerrOF3}. The energy level of $^{85}$Rb illustrated in Fig.~\ref{fig:Scheme}(a) is the basis of the nonlinear interaction. A weak laser (780.2 nm) in mode 1 tuned closely to the $5S_{1/2}\rightarrow5P_{3/2}$ transition induces a phase shift on a stronger laser (776 nm) in mode 2 tuned closely to the $5P_{3/2}\rightarrow5D_{5/2}$ transition. This nonlinear interaction can be characterized with the 3rd-order susceptibility $\chi^{(3)}\approx(N \mathfrak{d}_1 \mathfrak{d}_2)/(\epsilon_0\hbar^3\Delta^2_1\Delta_2)$, where $N$ is the atomic number density, $\mathfrak{d}_1$ and $\mathfrak{d}_2$ are the transition dipole moments for $5S_{1/2}\rightarrow5P_{3/2}$ and $5P_{3/2}\rightarrow5D_{5/2}$, respectively. The induced phase shift in mode 1 can be given by $\phi_1\sim (k_1n_2P_2L)/A$ with optimal detuning $\Delta_1\approx\Delta_2$, where $k_1$ is the wavenumber in mode 1, $n_2$ is the nonlinear refractive index, $P_2$ is the power of the beam in mode 2, $L$ is the length of the vapor-filled fiber, and $A$ is the optical mode area. 

Exemplary schemes with detailed descriptions are in Ref.~\cite{KerrOF2,KerrOF3}, which achieved the cross-phase shift up to $0.3$ mrad per photon with fast response time ($< 5$ ns) and low absorption ($<1\%$ ) at room temperature \cite{KerrOF2}. Moreover, cross-phase shift up to $\pi$ has been also demonstrated (e.g.~with $P_2\approx25\mu$W), as $\phi_1$ is proportional to $P_2$ (in the case without atomic saturation) \cite{KerrOF3}. Note that the shifted phase is guaranteed not to be induced from self-Kerr effects within this approach \cite{KerrOF2,KerrOF3}. Further improvements are expected by reducing the core diameter, increasing the atomic density, and extending $L$. It is suitable for the integration with fiber-optic communication network.

\subsection{Generating PSPs}

Suppose that coherent states $\ket{\sqrt{\mu}}_1\ket{\sqrt{\nu}}_2$ go through the two modes (1 for signal 2 for meter) of atomic vapor in HC-PCF illustrated in Fig.~\ref{fig:Scheme}(a). The interaction Hamiltonian can be written by $-\hbar\chi^{(3)}\hat{n}_1\hat{n}_2$ for time $t$ with the number operator $\hat{n}_i$ in $i$th mode. We assume that $\mu$ is small, while $\nu$ is large enough such that $\sqrt{\nu} \gtrsim 2\pi/\chi^{(3)} t$. Without loss of the generality, we set $d \equiv 2\pi/\chi^{(3)} t$ as an integer $\geq 2$, which is tunable by changing e.g.~the atomic density or length of HC-PCF. The state in the output modes is then written by 
\begin{eqnarray}
\begin{aligned}
&e^{\frac{2\pi i}{d}\hat{n}_1\hat{n}_2}\ket{\sqrt{\mu}}_1\ket{\sqrt{\nu}}_2\\
&=e^{\frac{2\pi i}{d}\hat{n}_1\hat{n}_2}\Big(\frac{1}{\sqrt{d}}\sum^{d-1}_{j=0}\sqrt{\frac{{\cal N}_{\mu,j}}{d}}\ket{j_d}\Big)_1\Big(\frac{1}{\sqrt{d}}\sum^{d-1}_{k=0}\ket{k_d}\Big)_2\\
&=\frac{1}{\sqrt{d}}\sum^{d-1}_{j=0}\sqrt{\frac{{\cal N}_{\mu,j}}{d}}\ket{j_d}_1\Big(\frac{1}{\sqrt{d}}\sum^{d-1}_{k=0}\omega^{jk}\ket{k_d}\Big)_2\\
&=\sum^{d-1}_{j=0}\frac{\sqrt{{\cal N}_{\mu,j}}}{d}\ket{j_d}_1\ket{\sqrt{\nu}\omega^j}_2.
\end{aligned}
\label{eq:genPSP}
\end{eqnarray}
A phase detection in mode 2 can identify $\ket{j_d}$ in mode 1:

(i) We can first consider a perfect discrimination of $\ket{j_d}$ by an ideal discrimination of the phase (i.e.~$j$) in mode 2. With the choice of $\sqrt{\nu}> d$, as the overlap between the coherent states $\ket{\sqrt{\nu}\omega^j}$ with different $j$ on a circle can be negligible, one can consider heterodyne measurements to fully discriminate $j\in\{0,1,...,d-1\}$. From Eq.~(\ref{eq:genPSP}), the probability of the generation of $\ket{j_d}$ is given by
\begin{equation}
\label{eq:probPNS}
P_{\mu,j}=\frac{{\cal N}_{\mu,j}}{d^2}=e^{-\mu}\sum^{\infty}_{n\in\mathbb{S}_j}\frac{\mu^{n}}{n!}.
\end{equation}

(ii) An alternative scheme to distinguish $\ket{1_d}$ only from $\ket{j(\neq1)_d}$ is illustrated in Fig.~\ref{fig:Scheme} (b) by using on-off detectors. Suppose that the state in mode 2 of Eq.~(\ref{eq:genPSP}) enters an input mode of the 50:50 beam splitter in Fig.~\ref{fig:Scheme} (b), where $\ket{\sqrt{\nu}\omega}$ enters the other input mode. As the beam splitter changes $\ket{\sqrt{\nu}\omega}\ket{\sqrt{\nu}\omega}\rightarrow\ket{\sqrt{2\nu}\omega}\ket{0}$, the event when only the upper detector clicks guarantees that the output state in mode 1 is $\ket{1_d}$, which we treat as a triggered event (others are non-triggered events). The probability of a triggered event is 
\begin{equation}
\label{eq:trigP}
\eta^{(t)}_{\nu,j}=\Big|\bracket{0}{\sqrt{\eta_{\rm det}\nu}(\omega^j-\omega)/\sqrt{2}}\Big|^2=\exp[-\eta_{\rm det}\nu|\omega^{j}-\omega|^2/4],
\end{equation}
where $\eta_{\rm det}$ is the detector efficiency, and the dark count is assumed to be negligible. The probability of the non-triggered event is $\eta^{(nt)}_{\nu,j}=1-\eta^{(t)}_{\nu,j}$. Note that $\eta^{(t)}_{\nu,1}=1$ and $\eta^{(nt)}_{\nu,1}=0$. On the other hand, $\eta^{(t)}_{\nu,j\neq1}$ tends to decrease as $\nu$ or $\eta_a$ increases, while it increases as $d$ increases. Note that low $\eta_{\rm det}$ can be compensated with the choice of large $\nu$.


\section{Application to secure quantum communications}
As PSPs can be directly applied to elementary quantum tasks, the tunable single-photon characteristics of PSPs with $\mu$ and $d$ can enrich the potential use in various quantum applications. However, here, we particularly consider a representative example showing the advantage of the single-photon characteristics from phase uncertainty: QKD with PSPs. 

\begin{figure}[b]
\centering
\includegraphics[width=3.4in]{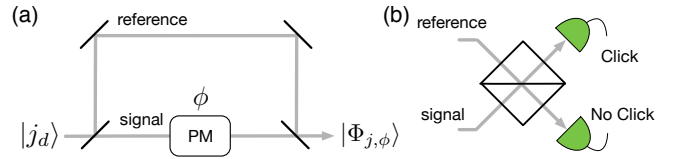}
\caption{(a) Schematic of the phase encoding QKD scheme on Alice's side with a Mach-Zehnder interferometer: $\ket{j_d}$ is separated into signal and reference pulses by a 50:50 beam splitter. A phase modulator (PM) encodes $\phi \in \{0, \pi/2, \pi, 3\pi/4\}$ into the relative phase between signal and reference pulses. The two pulses are recombined at the output. (b) $X$ or $Y$ basis measurement is performed on Bob's side with a 50:50 beam splitter and photon on-off detectors. The beam splitter is modified such that $\ket{\alpha}\ket{\beta}\rightarrow\ket{(\alpha+\beta)/\sqrt{2}}\ket{(\alpha-\beta)/\sqrt{2}}$ for the $X$ basis measurement and $\ket{\alpha}\ket{\beta}\rightarrow\ket{(\alpha+i\beta)/\sqrt{2}}\ket{(\alpha-i\beta)/\sqrt{2}}$ for the $Y$ basis measurement. 
}\label{fig:QKDScheme}
\end{figure}

\subsection{Bennett-Brassard 84 protocol}

Let us consider an implementation of the Bennett-Brassard 1984 (BB84) QKD protocol \cite{BB84} with PSPs. Assume that Alice wants to share a secret key with Bob. For this, Alice transmits quantum signals to Bob, encoding a logical bit (0 and 1) in one of two conjugated bases ($X$ and $Y$). Alice and Bob can detect any eavesdropper (Eve) attempting to collect the key information as such an activity causes a disturbance so that they can reject the key. Figure~\ref{fig:QKDScheme} shows the schematics of the encoding and decoding setups for the QKD with PSPs. 

Assume that $\ket{j_d}$ with mean photon number $2\mu$ is prepared on Alice's side and enters a phase encoding setup with a Mach-Zehnder interferometer in Fig.~\ref{fig:QKDScheme}(a). It is separated by the first 50:50 beam splitter into signal and reference pulses, and the phase modulator encodes $\phi \in \{0, \pi/2, \pi, 3\pi/4\}$ into their relative phase. Two pulses are recombined at the output of the interferometer as
\begin{equation}
\label{eq:afterPM}
\ket{\Phi_{j,\phi}}=\frac{1}{\sqrt{{\cal N}_{2\mu,j}}}\sum^{d-1}_{q=0}\omega^{-jq}\ket{\sqrt{\mu}\omega^q}_r\ket{\sqrt{\mu}\omega^{q+k}}_s,\\
\end{equation}
where the subscript $s$($r$) denotes the signal(reference) mode, and $k=\phi d/2\pi$ denotes the rotated phase for encoding. Notably, the randomness of the reference phase is guaranteed by the superposition of the states $\ket{\sqrt{\mu}\omega^q}$ with $q\in\{0,...,d-1\}$. This is contrast to the conventional QKD encoding using WCS in the form of $\ket{\sqrt{\mu}\omega^q}_r\ket{\sqrt{\mu}\omega^{q+k}}_s$, which requires an additional process for the phase randomization with an active phase modulator and a random number generator \cite{Cao2015}. The signal pulse is thus encoded in either $X$ or $Y$ basis, i.e,~ $\{\ket{0_x}_j,\ket{1_x}_j\}=\{\ket{\Phi_{j,0}},\ket{\Phi_{j,\pi}}\}$ or $\{\ket{0_y}_j,\ket{1_y}_j\}=\{\ket{\Phi_{j,\pi/2}},\ket{\Phi_{j,3\pi/4}}\}$. 

On Bob's side, a measurement is performed either in $X$ or $Y$ basis as illustrated in Fig.~\ref{fig:QKDScheme}(b). For $X$-basis measurement, the reference and signal states enter the input modes of a beam splitter such that
\begin{equation}
\nonumber
\begin{aligned}
\ket{0_x}_j&=\frac{1}{\sqrt{{\cal N}_{2\mu,j}}}\sum^{d-1}_{q=0}\omega^{-jq}\ket{\sqrt{\mu}\omega^q}_r\ket{\sqrt{\mu}\omega^{q}}_s\rightarrow\ket{j_d}\ket{0}\\
\ket{1_x}_j&=\frac{1}{\sqrt{{\cal N}_{2\mu,j}}}\sum^{d-1}_{q=0}\omega^{-jq}\ket{\sqrt{\mu}\omega^q}_r\ket{-\sqrt{\mu}\omega^{q}}_s\rightarrow\ket{0}\ket{j_d}.
\end{aligned}
\end{equation}
Thus, two logical encodings (either 0 or 1) can be discriminated by performing photon on-off detection at the two output modes. Similarly for the $Y$-basis measurement, two logical encoding can also be discriminated with a modulated beam splitter accordingly.

\begin{figure}
\centering
\includegraphics[width=3.0in]{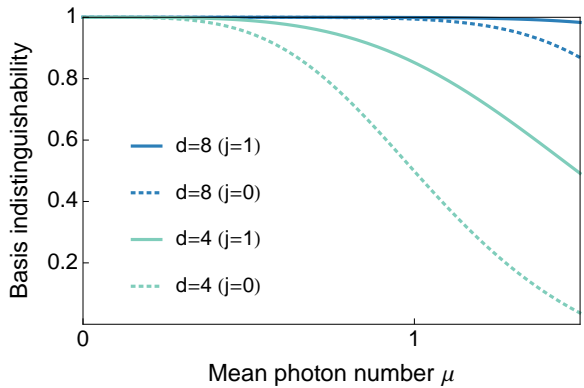}
\caption{The basis indistinguishabilities (between $X$ and $Y$ basis) are plotted by changing $\mu$ for $d=4,8$. The solid lines denotes the basis indistinguishabilities for $\ket{1_d}$, while the dotted lines are for $\ket{0_d}$. 
}\label{fig:FBIndep}
\end{figure}

An essential requirement to guarantee the security of the BB84 protocol is the basis indistinguishability so that Eve cannot distinguish the logical states encoded in two conjugate ($X$ and $Y$) bases. The basis indistinguishability can be investigated by the fidelity $\mathcal{F}_j(\rho_x,\rho_y)$ between two mixed states, $\rho_{x}=\ket{0_x}_j\bra{0_x}+\ket{1_x}_j\bra{1_x}$ and $\rho_{y}=\ket{0_y}_j\bra{0_y}+\ket{1_y}_j\bra{1_y}$ (the coefficient is omitted). Following the procedure in Ref.~\cite{Cao2015}, its minimum limit can be obtained as 
\begin{equation}
\label{eq:fidelity}
\mathcal{F}_j(\rho_x,\rho_y)\geq\frac{d\big|\sum^{d-1}_{q=0}\omega^{jq}e^{-2\mu+\mu\omega^{-q}}(e^{i\mu\omega^{-q}}+ie^{-i\mu\omega^{-q}})\big|}{\sqrt{2}{\cal N}_{2\mu,j}}.
\end{equation}
which is plotted in Fig.~\ref{fig:FBIndep} for $\ket{0_d}$ and $\ket{1_d}$ by changing $\mu$ and $d$. It shows that the encoded logical states in the two bases are more indistinguishable as $d$ increases or $\mu$ gets smaller. From the fact that $X$ and $Y$ basis encodings are indistinguishable both in $\ket{0_d}$ and $\ket{1_d}$ in the small-$\mu$ regime, we can see that a photon loss in the PSP is not significantly detrimental to the security of QKD.

\subsection{Key generation rate}

We analyze the key generation rate of QKD with PSPs. For WCSs, the key generation rate can be estimated based on the Gottesman-Lo-L\"utkenhaus-Preskill security analysis \cite{GLLP} as,
\begin{equation}
\label{eq:keyrate1}
R\geq -fQ_{\mu}H(E_{\mu})+Q_{\mu}\Omega\Big[1-H(E_{\mu}/\Omega)\Big],
\end{equation}
where $Q_{\mu}$ and $E_{\mu}$ are the overall gain and quantum bit error rate (QBER), respectively, for a given mean photon number $\mu$ of the signal state. Here, $f$ is the error correction efficiency, $H(p)=-p\log_2{p}-(1-p)\log_2(1-p)$ is the binary Shannon entropy function, and $\Omega$ is the fraction of Bob's detection orginating from single-photon signal emitted from Alice. We make here the pessimistic assumption that all errors come from single-photon states in the signal. For the non-decoy method, the fraction is estimated to be $\Omega=(Q_{\mu}-P_{\rm multi})/Q_{\mu}$, where $P_{\rm multi}$ denotes the probability of Alice's emitting a multi photon state as signal. On the other hand, by using the decoy state method \cite{decoy1,decoy2,decoy3}, the second term in Eq.~(\ref{eq:keyrate1}) is replaced with $Q_1[1-H(e_1)]$, where $Q_1=Y_1\mu e^{-\mu}$ is the gain of the single-photon component, $Y_n$ and $e_n$ are the yield and QBER of a given photon number $n$ respectively, and $\mu e^{-\mu}$ is the probability of generating a single-photon state in the Poission distribution. From the underlying assumption of the decoy state method \cite{decoy2}, i.e.~$Y_n({\rm decoy})=Y_n({\rm signal})$ and $e_n({\rm decoy})=e_n({\rm signal})$, Alice can vary $\mu$ for each signal. The variable $Y_n$ and $e_n$ can be deduced with the experimentally measured $Q_{\mu}$ and $E_{\mu}$. Therefore, in the decoy state method, Alice can pick up the mean photon number as $\mu=O(1)$ so that the key rate is obtained as $O(\eta)$ where $\eta$ is the overall transmission probability. On the other hand, the optimal $\mu$ for the non-decoy method is $\mu=O(\eta)$ yielding the key rate $R=O(\eta^2)$. As a result, the decoy method can substantially improve the key generation rate.

Let us estimate the key generation rate with PSPs. Suppose that Alice prepares a setup for generating $\ket{j_d}$ with probability $P_j$ in Eq.~(\ref{eq:probPNS}). The transmission probability of a single-photon state through an optical fiber is $\eta=10^{-0.21L/10}\eta_{\rm Bob}$ with distance $L$ and the efficiency of transmission and detection on Bob's side  $\eta_{\rm Bob}$. The transmission probability for $n$-photon signals is $\eta_n=1-(1-\eta)^n$. The yield is given by $Y_n=\eta_n+Y_0-\eta_nY_0=1-(1-Y_0)(1-\eta)^n$ where $Y_0$ is the dark count rate. The total gain and QBER are then obtained as $Q_{\mu}=Y_0+\sum^{d-1}_{j=0}\sum^{\infty}_{n\in\mathbb{S}_j}\{1-(1-\eta)^n\}e^{-\mu}\mu^n/n!=Y_0+1-e^{-\eta\mu}$ and $Q_{\mu}E_{\mu}=e_0Y_0+e_{\rm det}(1-e^{-\eta\mu})$, respectively, where $e_{\rm det}$ is the detection error rate. For the non-decoy method, the key generation rate of our scheme can be estimated by Eq.~(\ref{eq:keyrate1}) with
\begin{equation}
P_{\rm multi}=\sum^{d-1}_{j=2}P_{\mu,j}=e^{-\mu}\sum^{d-1}_{j=2}\sum^{\infty}_{n\in\mathbb{S}_j}\frac{\mu^{n}}{n!}.
\end{equation}

If $\ket{j_d}$ is identified with a measurement discriminating $j$ on Alice's side, we can employ decoy state methods: 

(i) For the case when $j$ can be fully identified by e.g.~heterodyne measurements with the choice of $\sqrt{\nu}>d$, it is possible for Alice and Bob to follow the passive decoy state method \cite{Mauerer2007,AYKI,PassiveDecoy}. One additional thing we should take care of is the fact that pseudo-number states have discrete phases so that the encoding with two bases X and Y are not perfectly indistinguishable. The key generation rate is then estimated as
\begin{equation}
\label{eq:keyrate3}
R\geq P_1Y_1\Big[1-fH(e^b_1)-H(e^p_1)\Big],
\end{equation}
with the pessimistic assumption that all errors occur with $\ket{1_d}$. Here, $\Delta_j=(1-F_j)/2Y_j$ is defined as the basis dependence with the fidelity in Eq.~(\ref{eq:fidelity}), and the bit and phase error rates are given respectively by \cite{Lo2007} $e^b_j=(e_0-e_{\rm det})(Y_0/Y_j)+e_{\rm det}$ and $e^p_j\leq e^b_j+4\Delta_j(1-\Delta_j)(1-2e^b_j)+4(1-2\Delta_j)\sqrt{\Delta_j(1-\Delta_j)e^b_j(1-e^b_j)}$.

(ii) We then consider the case when only PSPs can be discriminated from $\ket{j(\neq1)_d}$, e.g., by inefficient on-off detectors as illustrated in Fig.~\ref{fig:Scheme}(b). The gains and QBERs for triggered and non-triggered events are written by
\begin{equation}
\begin{aligned}
Q^{(t)}_{\mu}&=\sum^{d-1}_{j=0}Q^{(t)}_{\mu,\nu,j},~~E^{(t)}_{\mu}Q^{(t)}_{\mu}=\sum^{d-1}_{j=0}Q^{(t)}_{\mu,\nu,j}e_j,\\
Q^{(nt)}_{\mu}&=\sum^{d-1}_{j=0}Q^{(nt)}_{\mu,\nu,j},~~E^{(nt)}_{\mu}Q^{(nt)}_{\mu}=\sum^{d-1}_{j=0}Q^{(nt)}_{\mu,\nu,j}e_j.
\end{aligned}
\end{equation}
We define the total ratio $r\equiv Q^{(t)}_{\mu}/Q^{(nt)}_{\mu}$ and the ratio for $j\neq1$ as $r_{\nu,j\neq1}\equiv Q^{(t)}_{\mu,\nu,j}/Q^{(nt)}_{\mu,\nu,j}=\eta^{(t)}_{\nu,j}/\eta^{(nt)}_{\nu,j}$, which are experimentally known parameters from Eq.~(\ref{eq:trigP}). From the fact that $r_{\nu,2} \geq r_{\nu,j\geq2}$ and $r_{\nu,2} =r_{\nu,0}$, we can see that $Q^{(t)}_{\mu,\nu,j}\leq r_{\nu,0} Q^{(nt)}_{\mu,\nu,j}$ is always satisfied for $j\geq2$, which can be rewritten by $Q^{(t)}_{\mu}-Q^{(t)}_{\mu,\nu,0}-Q^{(t)}_{\mu,\nu,1}\leq r_{\nu,0}(Q^{(nt)}_{\mu}-Q^{(nt)}_{\mu,\nu,0})$. Since its left side is equal to $rQ^{(nt)}_{\mu}-r_{\nu,0}Q^{(nt)}_{\mu,\nu,0}-Q^{(t)}_{\mu,\nu,1}$, we arrive at $Q^{(t)}_{\mu,\nu,1}\geq (r-r_{\nu,0})Q^{(nt)}_{\mu}$.
The maximal QBER is also obtained as
\begin{equation}
\label{eq:QBERineq}
\begin{aligned}
e^b_1 &\leq \frac{E^{(t)}_{\mu}Q^{(t)}_{\mu}-e_0Q^{(t)}_{\mu,\nu,0}}{Q^{(t)}_{\mu,\nu,1}}\\
&=\frac{rE^{(t)}_{\mu}Q^{(nt)}_{\mu}-e_0r_{\nu,0}Q^{(nt)}_{\mu,\nu,0}}{Q^{(t)}_{\mu,\nu,1}}\leq \frac{rE^{(t)}_{\mu}Q^{(nt)}_{\mu}}{Q^{(t)}_{\mu,\nu,1}}.
\end{aligned}
\end{equation}
As a result, the key generation rate can be estimated by
\begin{equation}
\label{eq:keyrate4}
R\geq -fQ^{(t)}_{\mu}H(E^{(t)}_{\mu})+\min\bigg\{Q^{(t)}_{\mu,\nu,1}\Big[1-H(e^p_1)\Big]\bigg\}.
\end{equation}

\begin{figure}
\centering
\includegraphics[width=3.2in]{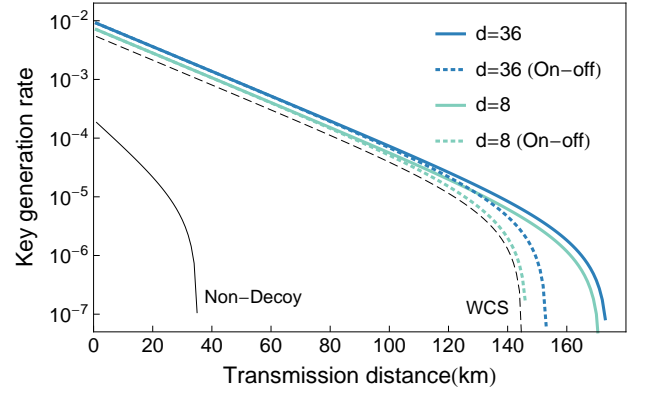}
\caption{The key generation rate of BB84 with PSPs. The thick solid lines are obtained when $j$ is fully discriminated, while the dotted lines are for the case when only PSP is discriminated by on-off detectors. 
The dashed line is obtained using phase randomized WCS. For our simulation, the experimental parameters in Ref.~\cite{GYS} are used: $f=1.16$, $\eta_{\rm det}=0.12$, $\eta_{\rm Bob}=0.045$, $Y_0=1.7\times10^{-6}$ and $e_{\rm det}=0.033$. The mean photon number in the signal is optimized as $\mu=1.0, 0.45, 0.08$ for $d=36, 8, 4$, respectively.
}\label{fig:KGRfigure}
\end{figure}

The key generation rates obtained by our scheme are plotted in Fig.~\ref{fig:KGRfigure}. Figure~\ref{fig:KGRfigure} shows that higher key generation rates can be achieved with PSPs over longer distances than a typical approach using phase randomized WCSs. For example, PSPs with $d\geq8$ allow us to obtain higher key generation rate than an active decoy state method using phase randomized WCSs, with photon detectors with low efficiencies ($\eta_{\rm det}=0.12$) as shown in Fig.~\ref{fig:KGRfigure}. We also note that QKD with PSPs does not require active phase modulation or a random number generator, which can be possible security loopholes with device imperfections. However, this result is obtained under the assumption of a fast response time of the PSP generator (as described in Sec.~\ref{sec:generation}). Therefore, the major challenge for the application of PSP to QKD is efficient generation of PSPs. 

\section{Remarks}

We have studied the single-photon characteristics of a superposed weak coherent state. This state, referred to as pseudo-single-photon state, mimics single photons with respect to both the number of photons and indeterminate phase. Not only is it close to the single-photon state with high fidelity but it also exhibits the fundamental behaviors of single photons such as antibunching and Hong-Ou-Mandel interference. The strength of the single-photon characteristics can be modulated by changing either the mean photon number $\mu$ or the number of phase $d$. This is in contrast to WCSs, which never exhibit such behaviors. Our result shows that the uncertainty between the number and phase indeed constitutes the fundamental characteristics of single photons. 

We have also presented a scheme to generate PSPs by employing, although not limited to, atomic vapor in HC-PCF \cite{KerrOF2,KerrOF3,KerrOF4,KerrOF5,KerrOF6,KerrOF7,KerrOF8}, which enables cross-phase modulation at room temperature with a fast response time. Despite recent progress in cross-phase modulation techniques in various systems \cite{CPM1,CPM2,CPM3,CPM4,CPM5,CPM6,CPM7}, they were mostly aimed at implementing quantum non-destructive measurements \cite{QND} or two-qubit gate operations \cite{CNOT1,CNOT2}, which are still demanding due to the requirement of either a strong nonlinearity or additional schemes with precise controls \cite{WCSforQC}. By contrast, a weak nonlinearity along with source or detector imperfections would suffice to implement our scheme. 

We have finally demonstrated that a QKD implementation with PSPs achieves higher key generation rates over longer distances than the typical QKD with phase randomized WCSs. This is a representative example showing the quantum advantage of quantum uncertainty over statistical randomization. 

Our work not only provides a fundamental understanding of single-photon characteristics but also opens an efficient way to explore them by modulating their strength. The tunable characteristics of PSPs may possibly enrich their potential applications. Exploring further fundamentals and applications of PSPs would be the next step of research.


\begin{thebibliography}{99}

\bibitem{noclon} W. Wootters and W. Zurek, Nature {\bf299} 802-803 (1982).

\bibitem{BB84}
C. H. Bennett and G. Brassard, \textit{Proceedings of IEEE International Conference on Computers. Systems, and Signal Processing}, 175-179 (IEEE Press, 1984).
\bibitem{Ekert91}
A. K. Ekert, Phys. Rev. Lett. {\bf 67}, 661 (1991).

\bibitem{Gisin2002}
N. Gisin, G. Ribordy, W. Tittel, and H. Zbinden, Rev. Mod. Phys. {\bf 74}, 145 (2002).
\bibitem{Scarani2009}
V. Scarani {\em et al.}, Rev. Mod. Phys. {\bf 81}, 1301 (2002).
\bibitem{Lo2014}
H.-K. Lo, M. Curty, and K. Tamaki, Nature Photonics {\bf8}, 595 (2014).

\bibitem{HOM}
C. K. Hong, Z. Y. Ou, and L. Mandel, Phys. Rev. Lett. {\bf 59}, 2044 (1987).

\bibitem{Teleportation}
C. H. Bennett,  G. Brassard, C. Crepeau, R. Jozsa, A. Peres, and W. K. Wootters, Phys. Rev. Lett. {\bf 70}, 1895  (1993).
\bibitem{DLCZ} L.-M. Duan, M. D. Lukin, J. I. Cirac, and P. Zoller, Nature {\bf414}, 413--418 (2001).
\bibitem{Pirandola2015} 
S. Pirandola, J. Eisert, C. Weedbrook, A. Furusawa, and S. L. Braunstein, Nature Photonics {\bf9}, 641--652 (2015).
 %

\bibitem{Eisaman2011}
M. D. Eisaman, J.  Fan, A. Migdall, and S. V. Polyakov, Review of Scientific Instruments {\bf82}, 071101 (2011).
\bibitem{Aharonovich2016}
I. Aharonovich, D. Englund, and M. Toth, Nature Photonics {\bf 10}, 631 (2016).
\bibitem{Schneider2018}
P.-I. Schneider {\em et al.} Optics Express {\bf26}, 8479 (2018).


\bibitem{decoy1}
W. Y. Hwang, Phys. Rev. Lett. {\bf 91}, 057901 (2003).
\bibitem{decoy2}
H.-K. Lo, X. Ma, and K. Chen, Phys. Rev. Lett. {\bf 94}, 230504 (2005).
\bibitem{decoy3}
X.-B. Wang, Phys. Rev. Lett. {\bf 94}, 230503 (2005).

\bibitem{Lo2007} 
H.-K. Lo and J. Preskill, Quantum Inf. Comput., {\bf7}, 0431 (2007).


\bibitem{SPbyWCS1}
X. Yuan, Z. Zhang, N. L\"utkenhaus, and X. Ma, Phys. Rev. A {\bf94} 062305 (2016)

\bibitem{SPbyWCS2}
P. Valente and A. Lezama, J. Opt. Soc. Am. B {\bf 34}, 924 (2017).

\bibitem{SPbyWCS3}
A. Aragoneses {\em et al.}, Optics Letters {\bf43}, 3806-3809 (2018).
\bibitem{SPbyWCS4} 
\'A. Navarrete, W. Wang, F. Xu, and M. Curty, New J. Phys. {\bf 20}, 043018 (2018).


\bibitem{KerrOF2}
V. Venkataraman, K. Saha, and A. L. Gaeta, Nature Photonics {\bf 7}, 138 (2012).
\bibitem{KerrOF3}
C. Perrella, P. S. Light, J. D. Anstie, F. Benabid, T. M. Stace, A. G. White, and A. N. Luiten, Phys. Rev. A {\bf88}, 013819 (2013).


\bibitem{old1} J. Janszky, P. Domokos, S. Szabo, and P. Adam, Phys. Rev. A {\bf51}, 4191 (1995).
\bibitem{old2} S. Szabo, P. Adam, J. Janszky, and P. Domokos, Phys. Rev. A {\bf53}, 2698 (1996).

\bibitem{Jkim2015}
J. Kim {\em et al.} Optics Communication {\bf337}, 79 (2015).

\bibitem{OddCat}
V. V. Dodonov, I. A. Malkin, and V. I. Man'ko, Physica {\bf72}, 597-615 (1974). 


\bibitem{Phoenix90} 
S. J. D. Phoenix,  Phys. Rev. A {\bf41}, 5132 (1990).

\bibitem{antib}
H. Paul, Rev. Mod. Phys. {\bf54}, 1061 (1982).

\bibitem{KSLee94} K. S. Lee, M. S, Kim, and V. B\v uzek, J. Opt. Soc. Am. B {\bf 11}, 1118-1129 (1994).

\bibitem{KerrOF4}
V. Venkataraman, K. Saha, P. Londero, and A. L. Gaeta, Phys. Rev. Lett {\bf 107}, 193902 (2011).

\bibitem{KerrOF5}
M. Bajcsy, S. Hofferberth, V. Balic, T. Peyronel, M. Hafezi, A. S. Zibrov, V. Vuleti\'c, and M. D. Lukin, Phys. Rev. Lett {\bf102}, 203902 (2009).

\bibitem{KerrOF6}
G. Epple {\em et al.}, Nature Communications {\bf5}, 4132 (2014).

\bibitem{KerrOF7}
M. R. Sprague {\em et al.}, Nature Photonics {\bf8} 287 (2014).

\bibitem{KerrOF8}
C. Perrella, P. S. Light, S. A. Vahid, F. Benabid, and A. N. Luiten, Physical Review Applied {\bf9}, 044001 (2018).

\bibitem{Cao2015} 
Z. Cao, Z. Zhang, H.-K. Lo, and X. Ma, New J. Phys. {\bf 17}, 053014 (2015).

\bibitem{GLLP}
D. Gottesman, H.-K. Lo, N. L\"utkenhaus, J. Preskill, Quant. Inf. Comput. {\bf 5} 325 (2004).

\bibitem{Mauerer2007} 
W. Mauerer and C. Silberhorn, Phys. Rev. A {\bf75}, 050305(R) (2007).
\bibitem{AYKI}
Y. Adachi, T. Yamamoto, M. Koashi, and N. Imoto, Phys. Rev. Lett. {\bf 99}, 180503 (2007).
\bibitem{PassiveDecoy}
X. Ma, and H.-K. Lo, New J. Phys. {\bf 10}, 073018 (2008).
\bibitem{GYS} 
C. Gobby, Z. L. Yuan, and A. J. Shields, Applied Physics Letters {\bf 84} 3762 (2004).

\bibitem{CPM7}
N. Matsuda, R. Shimizu, Y. Mitsumori, H. Kosaka, and K. Edamatsu, Nature Photonics {\bf 3}, 95 (2009).

\bibitem{CPM1}
O. Firstenberg {\em et al.} Nature {\bf502} 71 (2013).
\bibitem{CPM2}
I. -C. Hoi, {\em et al.}, Phys. Rev. Lett. {\bf 111}, 053601 (2013).
\bibitem{CPM3}
A. Feizpour, M. Hallaji, G. Dmochowski, and A. M. Steinberg, Nature Physics {\bf11} 905 (2015).
\bibitem{CPM4}
D. Tiarks, S. Schmidt, G. Rempe, and S. D\"urr, Science Advances {\bf 2}, e1600036  (2016). 
\bibitem{CPM5}
K. M. Becka, M. Hosseinia, Y. Duana, and V. Vuleti\'c, PNAS {\bf113} 9740 (2016).
\bibitem{CPM6}
Z. -Y. Liu, {\em et al.}, Phys. Rev. Lett. {\bf 117}, 203601 (2016).

\bibitem{QND}
W. J. Munro, K. Nemoto, R. G. Beausoleil, and T. P. Spiller, Phys. Rev. A {\bf71}, 033819 (2005).

\bibitem{CNOT1}
K. Nemoto and W. J. Munro, Phys. Rev. Lett. {\bf93}, 250502 (2004).
\bibitem{CNOT2}
W. J. Munro, K. Nemoto, and T. P. Spiller, New J. Phys. {\bf7}, 137 (2005).
\bibitem{WCSforQC}
H. Jeong, Phys. Rev. A {\bf73}, 052320 (2006).


\end{thebibliography}
\end{document}